\documentstyle{article}
\begin{document}
\title{Closed timelike curves in general relativity}
\author{W.B.Bonnor}
\maketitle
\setlength{\parindent}{0.5in}
\begin{abstract}
Many solutions of Einstein's field equations contain
closed timelike curves (CTC).  Some of these  solutions refer to
ordinary materials in situations which might occur in the laboratory, or
in astrophysics.  It is argued that, in default of a reasonable
interpretation of CTC, general relativity does not give a satisfactory
account of all phenomena within its terms of reference.

PACS number 0420
\end{abstract}

In general relativity a timelike curve in spacetime represents a possible
path of a physical object or an observer.  Normally such a curve will run
from past to future, but in some spacetimes
timelike curves can intersect themselves, giving a loop, or a {\em closed
timelike curve} (CTC).  CTCs suggest the possibility of time-travel with
its well-known paradoxes.

The first spacetime in which CTCs were noticed was that of G\"{o}del.
[1].  This represents a rotating universe without expansion, and requires
a negative cosmological constant.  As a model of physical reality it can
therefore be dismissed  because it is unlike the universe we live in.
Another simple spacetime containing CTCs is that of van Stockum [2]
which represents a cylinder of rigidly rotating dust; however, the cylinder
is of infinite length so it could not be realised in practice.

Since these early discoveries other spacetimes containing CTCs have been
found.  Nearly all of these have been regarded as of merely theoretical interest
because of some non-physical feature in their composition.
\footnote{An exception is the Kerr-Newman solution, which is asymptotically flat
and contains CTCs.  However, these are unlikely to be realised in astrophysics
or in the laboratory.}  Recently, however, there have been published
some solutions of Einstein's equations containing CTCs and representing
physical situations which in principle could be reproduced in the
laboratory, or might occur in astrophysics.

One such solution represents two spinning particles of masses $m_{1}, m_{2}$
and constant angular momenta $h_{1}, h_{2}$, their spins both parallel to their
line of separation.  The particles are
fixed on the $z$-axis at $z=\pm b$.
They are not supposed to represent black holes: they could be
copper spheres in a laboratory.  The set-up is axially symmetric and assumed independent of time.
so one may use the metric
\begin{equation}
ds^{2}=-f^{-1}[\exp \nu(dz^{2}+dr^{2})+r^{2}d\theta^{2}]+f(dt-wd\theta)^{2},
\end{equation}
$f,\nu,w$ being functions of $z$ and $r$ only.  The coordinates will be numbered
\[ x^{1}=z,\; x^{2}=r,\; x^{3}=\theta,\; x^{4}=t, \]
and their ranges are
\[-\infty <z<\infty,\; 0\leq r,\; 0\leq \theta \leq 2\pi,\; -\infty <t<\infty,\]
$\theta =0$ and $\theta=2\pi$ being identified.

Reference [3] contains details of the vacuum Einstein equations and an approximate
solution up to terms quadratic in the parameters $m_{1},m_{2},h_{1},h_{2}.$
One finds, as expected, that there is a singularity between the particles
representing a strut supporting them against their mutual gravitation.
However, the important expression for my purposes here is
\[ g_{33}=-f^{-1}[r^{2}-f^{2}w^{2}]. \]
If $\theta$ is a spacelike coordinate this should be negative, but it turns out
that on the axis of symmetry $r=0$ one can arrange this either between the
particles or outside them, but not both unless the parameters satisfy the
relationship
\begin{equation}
m_{1}h_{2}+m_{2}h_{1}=0.
\end{equation}

Let us suppose for the moment that (2) is not satisfied, and that there is a
region $D$ between the particles in which $g_{33}>0$.  Then $\theta$ is in $D$
a timelike coordinate which is cyclic because the hypersurfaces $\theta=0$
and $\theta=2\pi$ are identified.  Hence the spacetime contains CTC.

Eqn (2) compounds the mystery.  It can be written in the very simple form
\[ a_{1}+a_{2}=0, \]
where $a_{1}$ and $a_{2}$ are the angular momenta per unit mass of the two
particles.  There is no obvious reason why CTC should be absent in this case.

There are other simple physical systems in which CTC cannot be
avoided.  One consists of a static magnetic dipole $\mu$ and a static electric charge $e$
placed on dipole's axis.  They have masses $m_{1}$ and $m_{2}$ and they
need a strut between them to
counterbalance gravitation.  The set-up looks static, but in fact one
cannot solve the Einstein-Maxwell equations with an axially symmetric static
metric - one needs the stationary (1).  In [4] I found an approximate
solution for this system, correct to the quadratic terms in
the parameters  $m_{1},m_{2},e,\mu$.  Once again, positive values of $g_{33}$,
and therefore CTC,
must occur near the axis either between the particles or outside them.

These examples are approximate solutions of the field equations, but
somewhat similar exact solutions are known.  One arises from the
Perjes-Israel-Wilson (PIW) solutions [5][6] of the Einstein-Maxwell equations.
The PIW metrics are generalisations of the Papapetrou-Majumdar ones: they
represent sources consisting of spinning masses  bearing
electric charge and magnetic dipole moment, such that, in relativistic units,
the mass of each particle equals its charge, and its spin is equal to its
magnetic moment.  It was shown long ago [7] that two PIW particles in
axi-symmetric configuration engender CTC unless their parameters satisfy
a relation like (2).

Another example, shortly to be published [8], is an exact solution for a
finite spinning rod.  The solution can be constructed from the Papapetrou class
[9] which represents spinning massless objects.  It too has CTC.  One
can add mass as a perturbation to the exact solution and
the CTC persist.  CTC also exist in the vacuum spinning
C-metric [10].

These solutions refer to ordinary materials
that might occur in the laboratory, or in astrophysics.  They are
asymptotically flat.
What can CTC mean in these cases?  We surely cannot believe that, for
example, a charge and a magnet constitute a time machine!  I have suggested [3]
that, where it occurs between particles, a CTC region represents a
{\em torsion singularity} constraining their spin, but this explanation
does not seem to apply to the case of the spinning rod.

I believe there is an urgent need to find a convincing physical interpretation
of CTC.  They can no longer be
dismissed as curiosities occurring  in non-physical solutions.  We now know
that there are simple physical situations, such as that of the charge
and the magnet, within the terms of reference of general relativity, of which the theory
as currently understood does not give a satisfactory account.

\section*{References}
{[1]} G\"{o}del K 1949 {\em Rev. Mod. Phys.} {\bf 21} 447\\
{[2]} van Stockum W J 1937 {\em Proc. R. Soc. Edin.} {\bf 57} 135\\
{[3]} Bonnor W B 2001 {\em Class. Quantum Grav.} {\bf 18} 1381\\
{[4]} Bonnor W B 2001 {\em Phys. Letters A} {\bf 284} 81\\
{[5]} Perjes Z 1971 {\em Phys. Rev. Lett.} {\bf 27} 1668\\
{[6]} Israel W and Wilson G A 1972 {\em J. Math. Phys.} {\bf 13} 865\\
{[7]} Bonnor W B and Ward J P 1973 {\em Commun. Math. Phys.} {\bf 34} 123\\
{[8]} Bonnor W B 2002 {\em Class. Quantum, Grav.} in press\\
{[9]} Papapetrou A 1953 {\em Ann. Phys. Lpz.} {\bf 12} 309\\
{[10]}Bi\u{c}\'{a}k J and Pravda V 1999 {\em Phys. Rev.} {\bf 60} 044004
\end{document}